\documentstyle[aps,floats,graphicx]{revtex}

\begin{document}

\newcommand{\bec}{\begin{center}}
\newcommand{\ec}{\end{center}}
\newcommand{\be}{\begin{equation}}
\newcommand{\ee}{\end{equation}}
\newcommand{\beqn}{\begin{eqnarray}}
\newcommand{\eeqn}{\end{eqnarray}}
\newcommand{\bet}{\begin{table}}
\newcommand{\ent}{\end{table}}
\newcommand{\bib}{\bibitem}

\wideabs{

\title{
Strong mass effect on ion beam mixing in metal bilayers
}

\author{P. S\"ule, M. Menyh\'ard} 
  \address{Research Institute for Technical Physics and Material Science,\\
Konkoly Thege u. 29-33, Budapest, Hungary, sule@mfa.kfki.hu,www.mfa.kfki.hu/$\sim$sule\\
}

\date{\today}

\maketitle

\begin{abstract}
Molecular dynamics simulations have been used to study the mechanism of
ion beam mixing in metal bilayers.
We are able to explain the ion induced low-temperature phase stability and melting behavior of bilayers using only
a simple ballistic picture up to 10 keV ion energies.
The atomic mass ratio of the overlayer and the substrate constituents  
seems to be a key quantity in understanding atomic mixing.
The critical bilayer mass ratio of $\delta < 0.33$ is required for the
occurrence of a thermal spike (local melting) with a lifetime of $\tau > 0.3$ ps at low-energy
ion irradiation ($1$ keV) due to a ballistic mechanism.
The existing experimental data follow the same trend as the simulated values.
\\


\end{abstract}
}


 The mechanism of atomic transport in solids is a fundamental subject of material science
\cite{Allnatt}.
The atomic relocations, such as atomic mixing at interfaces under the effect of ion irradiation
has also been an important topic in the last decades \cite{Bellon,Okamoto}.
In these phenomena the atomic systems are driven far from equilibrium and the 
roughening of the interface acts in competition with the restoring forces of the relaxation
process.
In certain interfacial systems ion beam mixing (IM) or the elevated temperature seems to randomize the initially sharp interface
leading to amorphization \cite{Samwer,Gyulai}, to crystal-glass \cite{Okamoto} or liquid to glass transition \cite{Faupel}. In many bilayers no considerable IM occurs \cite{mix_exp1,mix_exp}.
As an explanation for the ion induced phase stability (instability) of interfaces, the thermal spike
(TS) model is widely accepted in the last two decades \cite{mix_exp,AverbackRubia}
which predicts the dependence of the mixing efficiency on thermodynamic quantities such as
heat of mixing.
Recently though we observed the presence of TS but
failed to find any effect of 
 heat of mixing on IM \cite{Sule1,Sule3}.
Thus accepting the importance of the TS we propose to understand its affect on IM as
a purely ballistic phenomenon \cite{Sule3}.


 In the present Letter we would like to show that
the atomic mass ratio has a dramatic effect on
IM in various metal bilayers. At 1 keV Ar$^+$ ion bombardment no real TS occurs above the atomic mass ratio of
$\delta=m_{overl}/m_{bulk}>0.33$ while considerable IM occurs if $\delta < 0.33$, where $m_{overl}$ and $m_{bulk}$ are the atomic masses in the overlayer and in the substrate (bulk).  
We will show that the measured mixing efficiencies follow the same trend as the simulated
IM.
We propose to understand IM as a ballistic  process and we find that the chemical
interdiffusion model \cite{Cheng} might be inappropriate.


Classical constant volume molecular dynamics simulations were used to simulate the ion-solid interaction
using the PARCAS code \cite{Nordlund_ref}.
The computer animations can be seen in our web page \cite{web}.
Here we only shortly summarize the most important aspects.
Its detailed description is given in \cite{Nordlund_ref} and details specific to the current systems in recent
communications \cite{Sule1,Sule2}.
We irradiate a series of bilayers (Al/Pt, Ti/Pt, Al/Ag, Cu/Pt, Cu/Au, Al/Cu, Ni/Ag, Ag/Au, Ti/Co, Ag/Ni, 
Cu/Ni) with 1 keV Ar$^+$ ions. In certain bilayers (Al/Pt, Ti/Pt, Al/Cu, Ni/Ag, Ti/Co) we 
increase the ion energy up to 10 keV in order to show that the mass effect
persists up to higher ion energies.
 The initial velocity direction of the
impacting atom was $7$ degrees with respect to the surface of the crystal (grazing angle of incidence).
  To obtain a representative {\em statistics},
the impact position of the incoming ion is varied
randomly within a $5 \times 5$ \hbox{\AA}$^2$ area.

 We used a tight binding many body potential given by Cleri and
Rosato (CR) \cite{CR}, to describe interatomic interactions.
We have chosen those bilayers for which atomic potentials are available.
This type of a potential gives a very good description of lattice vacancies, including migration
properties and a reasonable description of solid surfaces and melting \cite{CR}.
The bilayer AB cross potential is composed of the CR potentials of the A and B elements.
Taking the simple average of the elemental potentials the heat of mixing is set to $\Delta H_m \approx 0$.
This can be done since
 we have shown recently, that
we found no considerable dependence of IM on $\Delta H_m$ in Ti/Pt \cite{Sule1} up to
7 keV ion energy \cite{Sule3}.
The choice of $\Delta H_m \approx 0$ is also useful from that point of view that
we would like to show that even in the case of $\Delta H_m \approx 0$ (the lack of AB chemical bonding) considerable
IM occurs in certain bilayers.
In other bilayers the effect of $\Delta H_m$
on IM is also very weak if any \cite{Sule3}.
 The reliability of the AB crosspotential as an average of the elemental potentials is tested
in the case of CuAu, where an optimized CR potential is available for the alloy phase
\cite{CR2}. We find no considerable changes in the physical properties or in the equilibrium
structure. Although this does not prove the reliability of other crosspotentials, 
it suggests that the average potentials are very close to the optimized one.

  The  construction of the interface systems is given elsewhere \cite{Sule3}.
We only shortly summarize that
the interfaces have (111) orientation and the 
close packed directions are parallel.
\begin{figure}[hbtp]
\begin{center}
\includegraphics*[height=4.5cm,width=6cm,angle=0.0]{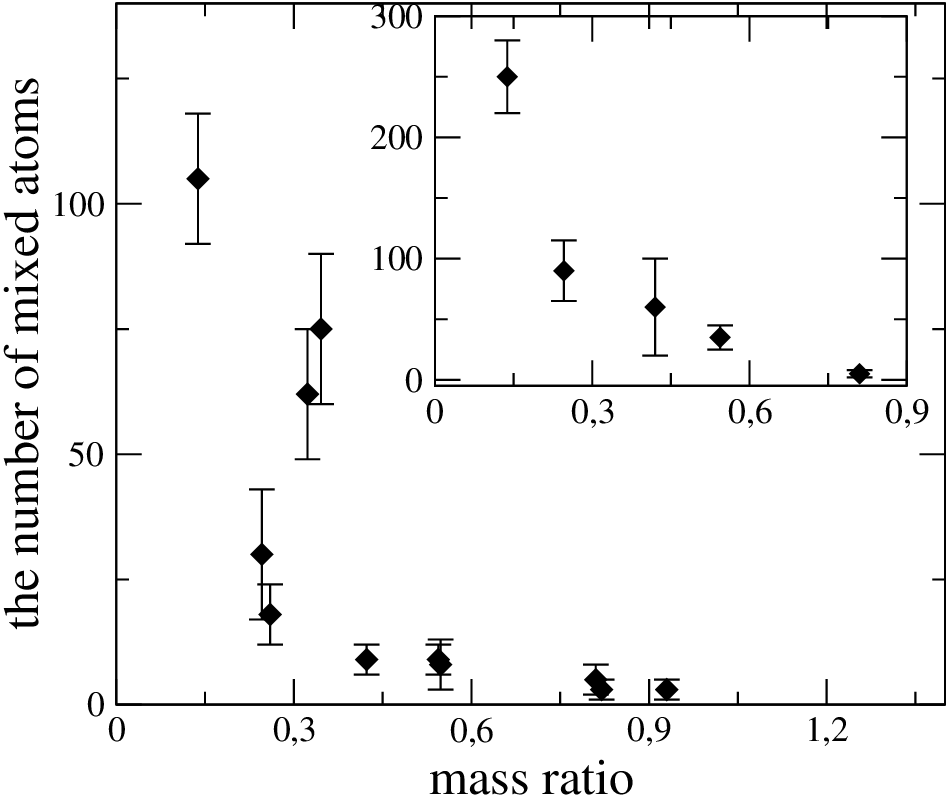}
\caption[]{
The number of intermixed atoms ($N_{mix}$, interfacial mixing) at various atomic mass ratio ($\delta$) in various
metal bilayers obtained at 1 keV Ar$^+$ irradiation.
Note the threshold $\delta \approx 0.33$ value below which
the bilayers exhibit an ion-beam mixing "catastrophe" and above which
only cascade mixing occurs.
The data points correspond from left to right to the bilayers Al/Pt (0.14), Al/Ag (0.24), Ti/Pt (
0.25), Cu/Au (0.32), Cu/Pt (0.33), Al/Cu (0.42), Ni/Ag (0.54), Au/Ag (1.83), Ti/Co (0.81), Cu/Ni (1.08) and 
Ag/Ni (1.82), respectively
(the atomic mass ratios are given in the paranthesis).
In those case where $\delta > 1$ we use $1/\delta$, because we find that $N_{mix}$ in AB and BA 
systems
are nearly equal.
The error bars denote standard deviations.
{\em Inset}: The number of mixed atoms is also shown for higher energies for
Al/Pt (6 keV), Ti/Pt (8 keV), Al/Cu (10 keV), Ni/Ag (9 keV) and for Ti/Co (10 keV) as a function
of the mass ratio.
}
\label{mass}
\end{center}
\end{figure}
The thickness of the upper layer is 4-8 monolayer (ML), while the bulk is constructed
from 36 MLs in those samples which subjected to 1 keV irradiation.
These samples includes roughly 45000 atoms.
At higher irradiation energy we put a thicker overlayer with 8-16 MLs and a substrate
with 90 MLs (550000 atoms).
The interfacial systems 
are created as follows:
the overlayer is put by hand on the (111) substrate (bulk) and various structures are probed
and are put together randomly. Finally that one is selected which has the smallest
misfit strain prior to the relaxation run.
The remaining misfit is properly minimized during the relaxation
process so that the overlayer and the substrate layers keep their original crystal structure and we get an
atomically sharp interface.

  We carry out {\em liquid and hot atom analysis} in order to study the role of highly mobile particles
in IM \cite{Sule3}.
To do so we need to calculate the
time dependent temperature of the $i$th high energy particle ($T_{i,local}(t)$) which can be given as follows:
$\frac{1}{2} m_i v_i^2(t) = \frac{3}{2} k T_{i,local}(t)$
where $m_i$ and $v_i$ are the atomic mass and the velocity of the $i$th high energy particle and $k$ is the Boltzmann constant.
We allow
some temperature fluctuation for the liquid atoms above $T_m$ (the melting point), hence for a hot atom and for recoils we use the arbitrary definition
$T_m+1000 < T_{local}$  \cite{Sule3}.
Hot atoms are those energetic particles which have a medium long mean path
($\sim 10$ \hbox{\AA})
with an average kinetic energy of 0.1-1 eV.
\begin{figure}[hbtp]
\begin{center}
\includegraphics*[height=4.5cm,width=6cm,angle=0.0]{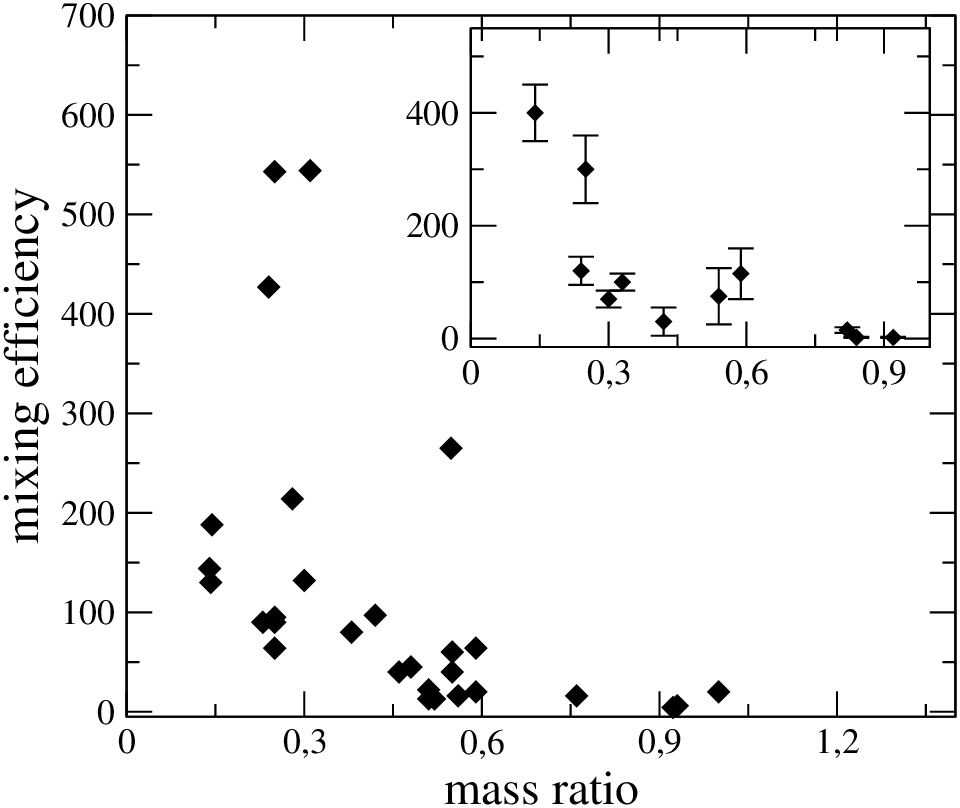}
\caption[]{
The low-temperature experimental mixing efficiency (\hbox{\AA}$^5$/eV) as a function of the
atomic mass ratio ($\delta$). The values are taken from refs. \cite{mix_exp1,mix_exp}.
In those cases where $\delta > 1$ we use instead $1/\delta$ because we find that there is no
serious difference in mixing between AB and BA bilayers.
{\em Inset}: The calculated (simulated) $\xi_{IM}$ (\hbox{\AA}$^5$/eV) vs. $\delta$.
}
\label{exp_mass}
\end{center}
\end{figure}
  Liquid atoms have shorter, recoils have much longer mean free path.
A liquid atom can be considered as a nearly standing atom during the lifetime of the TS and only vibrationally
excited around its equilibrium position while a hot atom jumps to another position with
a velocity of $v_i \approx 1$ \hbox{\AA}/ps \cite{Sule3}. 
The hot atoms are present till the end of the TS hence cascade and TS
mixing coexists in certain interfacial systems.
The more precise definition of a hot atom is given elsewhere \cite{Sule3}.

 We calculate the number of mixed atoms ($N_{mix}$) and the simulated mixing efficiency $\xi_{IM}^{sim}=\frac{\langle R^2 \rangle}{6 n_0 E_{D_N}}$, where $\langle R^2 \rangle$, $n_0$ and $E_{D_N}$, 
are the calculated mean square atomic displacement through the interface per atom, the atomic density in the
upper layer and the deposited nuclear energy.
We exclude from $\langle R^2 \rangle$ atomic displacements which do not lead to
broadening at the interface (self-atomic mixing) because the experimental $\xi_{IM}$ is
calculated from broadening at the interface \cite{mix_exp,AverbackRubia}.
Further calculational details are given in ref. \cite{NordlundNIMB00}.

 We investigate the influence of the atomic mass ratio on ion beam mixing.
We ion bombarded various bilayer systems with different atomic mass ratios and 
find that below a threshold ratio ($\delta < 0.33$) the magnitude of
intermixing ($N_{mix}$) is enhanced abruptly.
The results are summarized in FIGs (1)-(2).
In FIG (1) we plot the simulated $N_{mix}$ vs. $\delta$ and 
the experimental $\xi_{IM}$ vs. $\delta$ in FIG (2) collected from refs. \cite{mix_exp1,mix_exp}.
In the inset FIG (2) we also give the simulated $\xi_{IM}$ vs. $\delta$
for bilayers for which high energy (up to 10 keV) simulations are available and also for
the other bilayers for which $\xi_{IM}^{sim}$ is calculated at 1 keV ion energy.
In FIG (2) wee see that the increase in $\xi_{IM}$ occurs between $\delta \approx 1/3$
and $1/2$ which is rather similar to the inset FIG (1).
\begin{figure}[hbtp]
\begin{center}
\includegraphics*[height=4.5cm,width=6cm,angle=0.0]{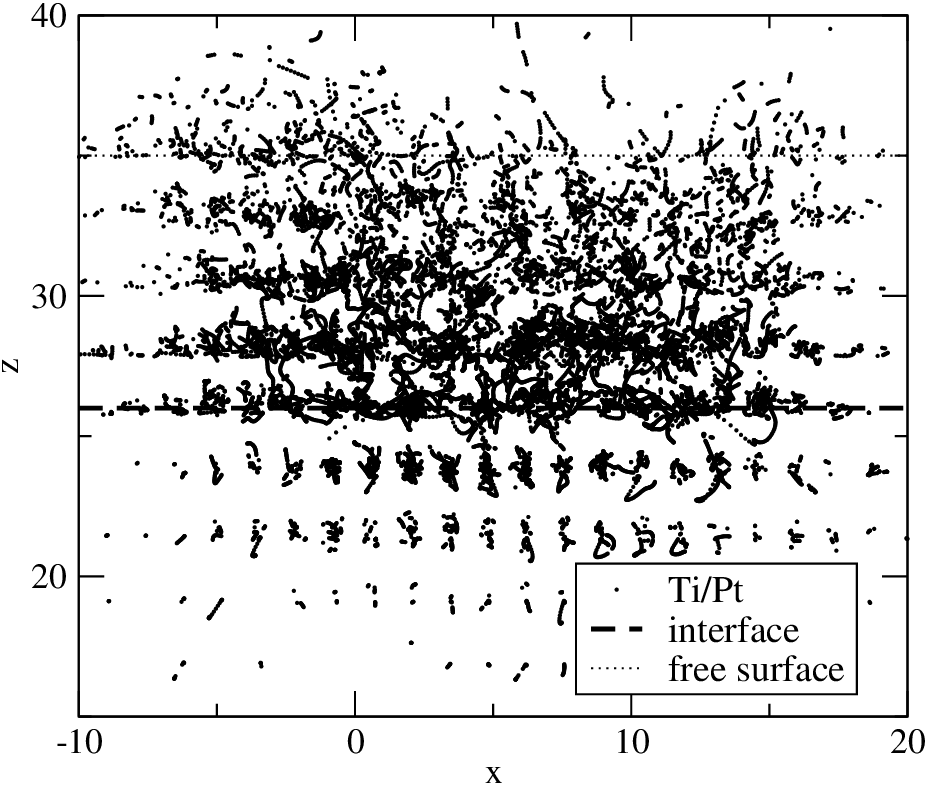}
\caption[]{
The $xz$ atomic positions of the recoils and the hot atoms at 1 keV ion energy collected up to $4
$ ps in Ti/Pt.
The z-coordinate is the depth position.
The dashed line denotes the interface.
}
\label{cascade1}
\end{center}
\end{figure}
 In FIG ~(\ref{exp_mass}) the experimental mixing efficiency values are obtained as follows \cite{mix_exp1}: ($\xi_{IM} \approx \frac{D t}{\Phi F_D}$), where $D t$ is the diffusion length, $\Phi$ is the fluence and $F_D$ is the deposited energy at the interface). 
Although there is some scatter in the data
at around $\delta \approx 1/3$, $\xi_{IM}$ increases heavily in accordance with the
$N_{mix}$ values shown in FIG ~(\ref{mass}).
Although the values in FIG ~(2) are obtained at high energies, however,
they can be compared with our $\xi_{IM}^{sim}$ values since IM occurs
primarily in the subcascade region, where the energy is in the range we applied \cite{NordlundNIMB00}.

  In particular in Al/Pt we find a strong amorphization and broadening at the interface
which is in accordance with measurements \cite{Samwer,Gyulai,Sule3,Hung}.
In Cu/Pt and in Cu/Au the phase stability of the interface is also very weak
in accordance with ion irradiation experiments \cite{Hung}.
In Cu/Au 1 MeV ion bombardment results in strong broadening \cite{mix_exp,Hung}.
In bilayers Al/Ag and in Au/Ag we find a relatively weak interfacial mixing ($N_{mix} < 30$),
however, $\xi_{IM}^{sim}$ is large ($\xi_{IM}^{sim} \approx 300 \pm 120$ in Al/Ag,
$\xi_{IM}^{sim} \approx 115 \pm 45$ in Au/Ag).
A large measured value of $\xi_{IM}^{exp} approx 265$  is found in Au/Ag 
\cite{mix_exp1} and this data point is plotted at $1/\delta=0.55$ in FIG (2).
Interestingly the atomic mobility is relatively large in these systems while the gross number of
mixed atoms ($N_{mix}$) is small. 
We attribute this "anomalous" behavior of Au/Ag to the tendency of crater formation in Au \cite{AverbackRubia}.
It has been shown, recently, 
that crater formation enhances mass transport hence atomic mobility between the interface and the free surface \cite{Sule2}.

 In FIG ~(\ref{exp_mass}) the experimental mixing efficiency values \cite{mix_exp1,mix_exp} ($\xi_{IM} \approx \frac{D t}{\Phi F_D}$), where $D t$ is the diffusion length, $\Phi$ is the fluence and $F_D$ is the deposited energy at the interface) are plotted against
the atomic mass ratio.
Although there is some scatter in the data
at around $\delta \approx 1/3$, $\xi_{IM}$ increases heavily in accordance with the
$N_{mix}$ values shown in FIG ~(\ref{mass}).
Although the values in FIG ~(2) are obtained at high energies, however,
they can be compared with our $\xi_{IM}^{sim}$ values since IM occurs
primarily in the subcascade region, where the energy is in the range we applied \cite{NordlundNIMB00}.

  For bilayers with $\delta > 0.33$ starting with Al/Cu ($\delta = 0.42$) we got a weak interfacial mixing in accordance
with the MD \cite{Colla,NordlundPRB99} and experimental studies \cite{mix_exp1}.
The strong mass effect can be understood as the ballistic mechanism 
plays an essential role in IM and
may stem
from an increased backscattering of light overlayer atoms from the
heavy substrate \cite{Sule3}.
\begin{figure}[hbtp]
\begin{center}
\includegraphics*[height=4cm,width=5.5cm,angle=0.0]{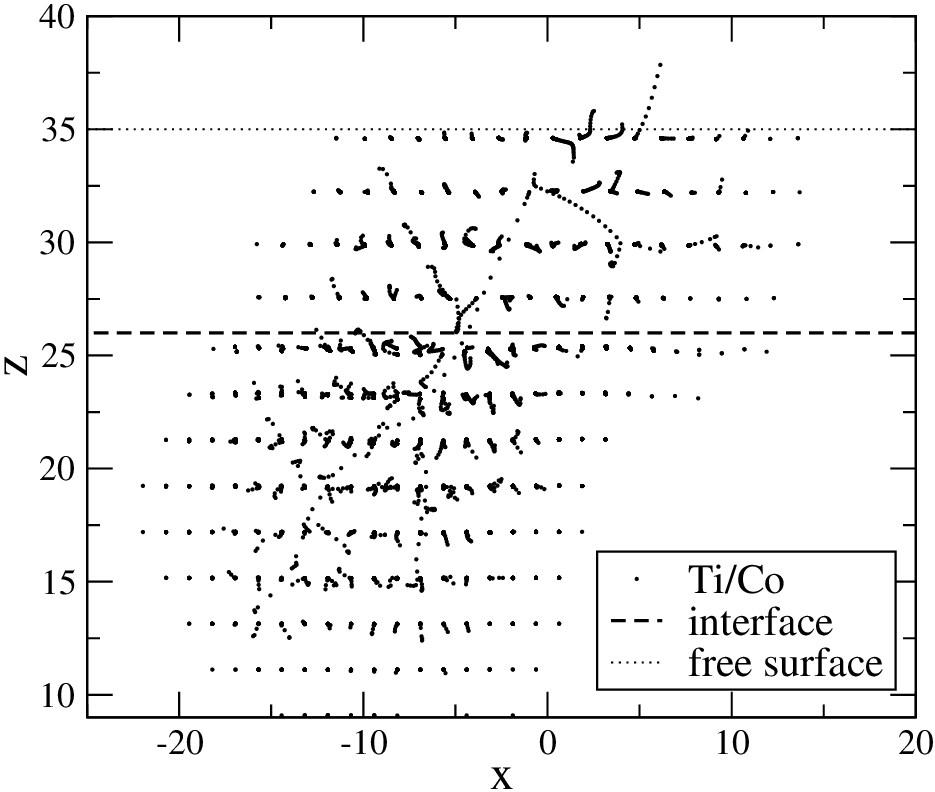}
\caption[]{
The $xz$ positions of the recoils and the hot atoms at 1 keV ion energy collected up to $0.3$ ps in Ti/Co.
The z-coordinate is the depth position.
The dashed line denotes the interface.
}
\label{cascade2}
\end{center}
\end{figure}
Moreover the density of the collisional cascades depends on the bilayer mass ratio \cite{Sule3}.
It should also be stressed that in pure elements or even in metal
alloys we find a much weaker IM and shorter TS \cite{web,Sule2} due to
the strong (re)ordering forces.
 In order to elucidate the relative roles of ballistics and TS in the
interfacial mixing 
 we examined the prototypical bilayer Ni/Ag ($\delta \approx 0.54$) and
found a very weak interfacial mixing at 1 keV bombardment and no occurrence of a real TS period. 
The energetic atoms (recoils, hot atoms) dissappear at less then 0.3 ps,
which is typically the end of the cascade period.
This system is relatively well studied
theoretically \cite{Colla} at 10 keV ion energy and being a typical
example of a segregating system which has a high positive heat of mixing 
in the liquid \cite{Colla}.
At 9 keV we find a real thermal spike which persists up to 3 ps.
The number of mixed atoms ($N_{mix}$) is, however, much smaller then in Al/Pt or in Ti/Pt (inset FIG ~\ref{mass})).
Therefore, although there is an increase in $N_{mix}$, the trend remains the same: the effect of
the mass ratio is also robust at higher energies.
We attribute, however, the weak IM in Ni/Ag not to the positive heat of mixing, but to the weak mass
effect in this system ($0.33 < \delta \approx 0.54$). 
In Ti/Co we find no TS up to 10 keV as well as IM is very weak in this system ($\delta \approx 0.82$).
  It should be emphasized that the similar situation is true for all the bilayer samples
which have $\delta > 0.33$ (FIG ~(\ref{mass})).
This observation clearly indicates a robust mass effect when $\delta < 0.33$, hence a kinematic and ballistic
picture seems to be sufficient for describing IM in bilayer systems.
In those systems, where TS does not occur, cascade mixing is the only IM effect.

  The cascade events during the ballistic period is shown in FIGs ~(\ref{cascade1})-
(\ref{cascade2}).
In the case of Ti/Pt we get a dense collision cascade, the recoils (hot atoms) are concentrated 
within a
smaller region due to couple of reflections while in Ti/Co the high energy particles
are scattered in a larger volume hence the deposited energy spreads over a larger irradiated
region. Therefore if the mass ratio $\delta > 0.33$, the cooling of the cascade
is ultrafast due to the low concentration of the recoils (hot atoms).
Indeed, we found that the average atomic concentration of the hot atoms in the irradiated zone ($V \approx 1000 \hbox{\AA}^3$) is
around $10^{22}$ atom/cm$^3$ for $\delta > 0.33$ and $4 \times 10^{22}$ atom/cm$^3$ for $\delta <
 0.33$.
These values should also be compared with the average atomic concentration
of $5-8 \times 10^{22}$ atom/cm$^3$ in metals.
If the mass ratio drops below $0.33$, the hot particles are still present in the TS.
One can see that for $\delta < 0.33$ the peak hot atom concentration is close to the 
atomic concentration. In these systems we no longer have a simple liquid ensemble, it is rather
a superheated system \cite{Sule3}.
The mass effect is even more pronounced at higher energies. The corresponding plots of
the collisional cascades in Al/Pt at 6 keV ion energy can be seen in a web page \cite{web2}.
We have also shown that the density of the cascade is strongly affected by $\delta$ \cite{Sule3,web}.

 {\em Qualitatively we explain the observed strong mass effect as follows:}
In elastic collisions of the recoils (energetic light particles from the overlayer) with the heavier substrate atoms
the kinetic energy of the moving atoms is partly transferred to the heavier atoms, which
, however, might not be kicked out of their positions because of the large mass difference.
The colliding heavy partner of the recoil becomes vibrationally excited, which means that its rms amplitude
of thermal vibrations becomes equal to about 50 \% of the interatomic distance. That is basically
the Lindeman`s criterion for lattice instability during melting: a crystal melts when the rms thermal
displacement of atoms from their equilibrium positions become large enough to invade their
nearest-neighbor spaces \cite{Okamoto}.
For such thermal displacements, the thermal expansion would far exceed the critical value for
shear instability (the Born criterion, the lost of at least one of the shear moduli) leading to mechanical instability \cite{Okamoto}.
The neighborhood of this hot heavy atom is heated up and local melting
occurs (TS).
  When the mass of a recoil and the colliding partner is comparable the target atom
might be displaced from its original position leaving a vacant site. In this case
the slowing down of the recoil in the bulk does not result in local melting because the
kinetic energy of the recoil spreads over a too large volume.

  The important question remained to be answered what is the reason of the critical mass ratio
of $\delta \approx 0.33$?
We do believe that the threshold value is due to the emergence of a strong backscattering of the recoiling light atoms at the interface. Below
this value hence the energy deposition becomes extremely effective at the interface through energy
transfer to the standing heavy atoms.
The backscattering effect at the interface results in the confinement of the light recoils
in the overlayer which leads to superheating. This is the primary reason of the high concentration
of hot atoms in these bilayers (see FIGs ~(\ref{cascade1})).
The interfacial backscattering phenomenon can be attributed partly to the mass difference
and also to other effects such as the difference in the cohesive energies in the
substrate and in the overlayer \cite{Sule3}.

 In summary, we have shown that intermixing in metal bilayers strongly depends
on the relative masses of the constituents under the effect of ion irradiation.
There exists a threshold mass ratio value below which the interface system is unstable
against ion bombardment. 
We propose to understand ion beam mixing as a ballistic process.
The observed strong mass effect in heterophases might be an important topic in preparation
of thin films and multilayers especially with great technological importance.

{\small
This work is supported by the OTKA grant F037710
from the Hungarian Academy of Sciences. We greatly
acknowledge conversations with K. Nordlund.}


\end{document}